\documentclass[]{spie}  
\usepackage[]{graphicx}
\usepackage{amsmath} 
\usepackage{amsfonts}
\usepackage{mathrsfs}
\usepackage{comment}
\title{Multi-modality super-resolution loss for GAN-based super-resolution of clinical CT images using \\micro CT image database} 

\author{Tong ZHENG\supit{a}, Hirohisa ODA\supit{a}, Takayasu MORIYA\supit{a}, Takaaki SUGINO\supit{a}, \\Shota NAKAMURA\supit{b}, Masahiro ODA\supit{a}, Masaki MORI\supit{c}, Horitsugu TAKABATAKE\supit{d}, \\Hiroshi NATORI\supit{e} and Kensaku MORI\supit{a,f,g}
\skiplinehalf
\supit{a}Graduate School of Informatics, Nagoya University, Japan;\\
\supit{b}Nagoya University Graduate School of Medicine, Japan;\\ 
\supit{c}Sapporo-Kosei General Hospital, Japan;\\
\supit{d}Sapporo Minami-sanjo Hospital, Japan;\\
\supit{e}Keiwakai Nishioka Hospital, Japan; \\
\supit{f}Information Technology Center, Nagoya University, Japan; \\
\supit{g}Research Center of Medical Bigdata, National Institute of Informatics, Japan
}

\authorinfo{Further author information: (Send correspondence to Tong Zheng or Kensaku Mori) \\ Tong Zheng.: E-mail: tzheng@mori.m.is.nagoya-u.ac.jp \qquad Kensaku Mori.: E-mail: kensaku@is.nagoya-u.ac.jp} 

\begin{document}
  \maketitle 

\begin{abstract}
This paper newly introduces {\it multi-modality loss function} for GAN-based super-resolution that can maintain image structure and intensity on unpaired training dataset of clinical CT and micro CT volumes. Precise non-invasive diagnosis of lung cancer mainly utilizes 3D multidetector computed-tomography (CT) data. On the other hand, we can take $\mu$CT images of resected lung specimen in 50 $\mu$m or higher resolution. However, $\mu$CT scanning cannot be applied to living human imaging. For obtaining highly detailed information such as cancer invasion area from pre-operative clinical CT volumes of lung cancer patients, super-resolution (SR) of clinical CT volumes to $\mu$CT level might be one of substitutive solutions. While most SR methods require paired low- and high-resolution images for training, it is infeasible to obtain precisely paired clinical CT and $\mu$CT volumes. We aim to propose unpaired SR approaches for clincial CT using micro CT images based on unpaired image translation methods such as CycleGAN or UNIT. Since clinical CT and $\mu$CT are very different in structure and intensity, direct appliation of GAN-based unpaired image translation methods in super-resolution tends to generate arbitrary images. Aiming to solve this problem, we propose new loss function called {\it multi-modality loss function} to maintain the similarity of input images and corresponding output images in super-resolution task. Experimental results demonstrated that the newly proposed loss function made CycleGAN and UNIT to successfully perform SR of clinical CT images of lung cancer patients into $\mu$CT level resolution, while original CycleGAN and UNIT failed in super-resolution.
\end{abstract}



\section{INTRODUCTION}
Lung cancer causes the largest number of deaths per year among cancers of male in Japan\cite{Cancerdeath}. Precise non-invasive diagnosis of lung cancer mainly uses clinical CT images. For more precise clinical diagnosis including diagnosing cancer invasion areas, super-resolution (SR) of clinical CT image to $\mu$CT image resolution level would be one of options. Most SR methods usually require paired training dataset. However, it is infeasible to collect paired clinical and $\mu$CT volumes. 

Unsupervised SR methods that do not require paired LR and HR images are very few. Most of these approaches are derived from image translation methods. One of them is CinCGAN\cite{yuan2018unsupervised}. But loss terms proposed in this paper did not consider relevance of input LR images and the output HR images, a key factor in medical image processing. Zhao et al.\cite{Zhao2018UnsupervisedDL} replaced generators of CycleGAN to achieve SR. But this method did not introduce any novel loss terms and even gave up the identity loss in conventional CycleGAN\cite{CycleGAN}, which may cause serious deformation of SR image, making it very different with the input LR imgae. Ravì et al.\cite{ravi2019adversarial} proposed an unsupervised image SR method for endomicroscopy. However, this method requires the fiber positions in specific imaging devices; since CT volumes are shot with different devices, we could not adapt this method directly to CT volumes. Inspired by these approaches, we aim to use unpaired image translation approaches like CycleGAN\cite{CycleGAN} or UNIT\cite{UNIT} for super-resolution of clinical CT. However, the original loss function of CycleGAN and UNIT was not designed to maintain similarity of input images and corresponding output SR images. This drawback makes CycleGAN and UNIT tend to generate arbitrary images in SR. It is important to design a loss function that can maintain the similarity of input images and corresponding output images.

This paper proposes unsupervised SR approaches for SR of clinical CT images into $\mu$CT scale. No paired clinical CT and $\mu$CT volumes are required for training. We introduce a novel loss function for preventing deformation and change of intensity distribution from original domain (clinical CT) but performs nice SR into $\mu$CT-level. Further, we introduce network structures of generative models based on CycleGAN or UNIT. We modify them for SR as SR-CycleGAN and SR-UNIT, respectively. Novel loss function allow us to perform SR of lung tissues on clinical CT volumes.

Contributions of this paper are summarized as follows:  1) novel loss function for unpaired dataset, 2) novel network structures for SR and 3) application to SR of clinical CT volumes of human lung tissues into $\mu$CT-level.

   \begin{figure}
   \begin{minipage}{1.0\hsize}
    			\begin{center}
    				\includegraphics[width=0.99\textwidth]{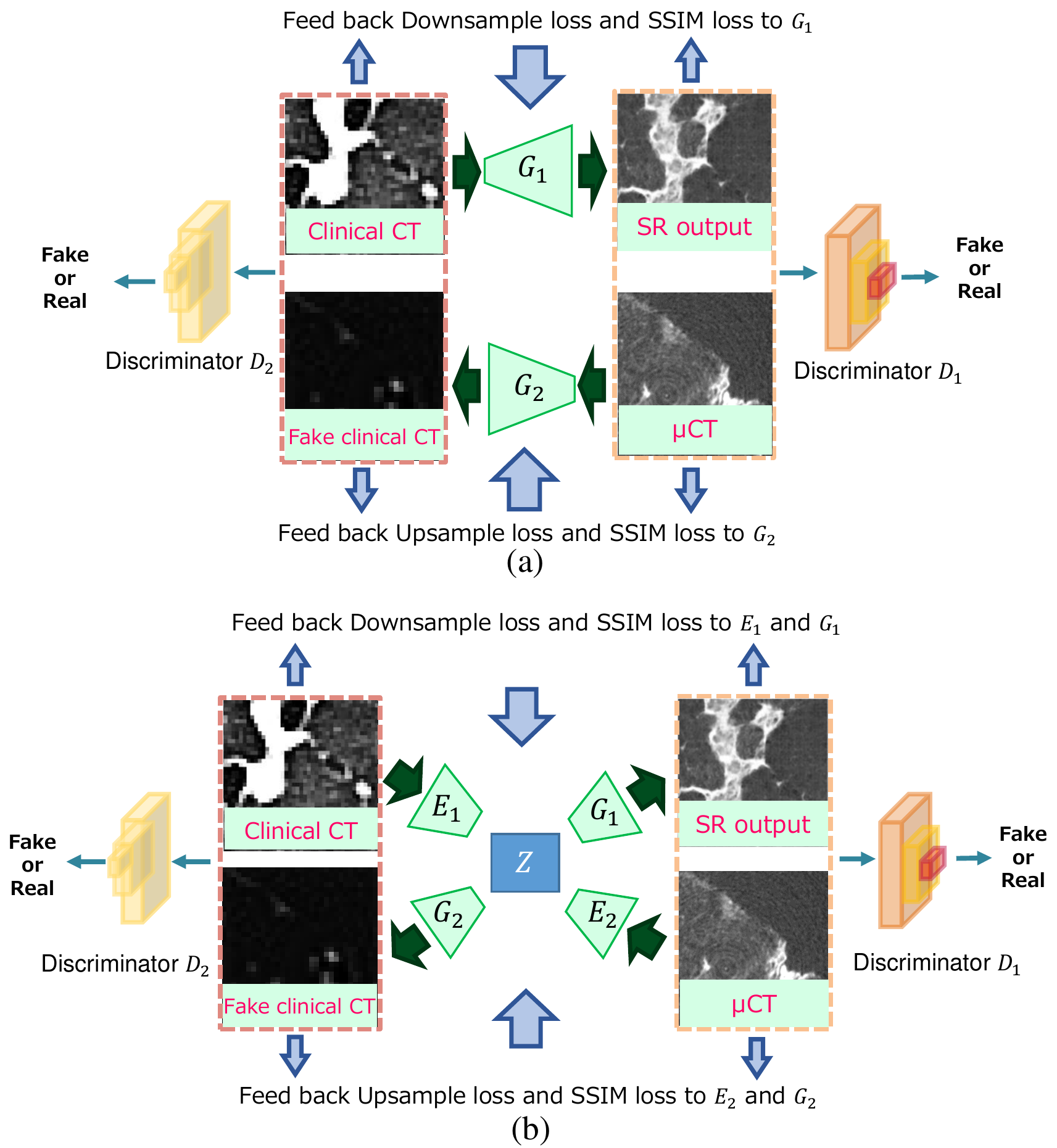}
    			\end{center}
    \end{minipage}
    	\\
   \caption[example] 
   { \label{fig_SRCycleGAN_AND_UNIT} 
Network structures of (a) SR-CycleGAN and (b) SR-UNIT. Modification to both network is same: replacement of generator $G_1$ with a Resnet\cite{he2016deep}-based SR network, replaced generator $G_2$ with network of which the length and width of output image size is 1/8 of input. We also added loss terms named as ``multi-modality SR loss" during training phase. Modified network structure and newly proposed loss functions made both SR-CycleGAN and SR-UNIT successfully performed SR of clinical CT to $\mu$CT scale.}
   \end{figure}

\section{Methods}
\subsection{Overview}
We propose the loss function named {\it multi-modality loss function} for GAN-based super-resolution on  unpaired dataset. We evaluate the effectiveness of proposed loss function by implementing it with CycleGAN or UNIT. We compare the modified models with original CycleGAN or UNIT.

Network training using clinical CT and $\mu$CT volumes is required. We assume that they have around 8-times difference in resolution. We train our network using 2D patches cropped from clinical or $\mu$CT volumes. We set the patch sizes from clinical and $\mu$CT volumes are 32$\times$32 pixels and 256$\times$256 pixels. Clinical CT and $\mu$CT images of same patients are used for network training.

\subsection{Multi-modality super-resolution loss (MMSR Loss) }

Since CycleGAN and UNIT are designed for domain translation, such as Monet's paintings to Gogh's ones, they does not guarantee the generated images are similar to the original images. Regardless of SR, we would like to keep the structure similarity on clinical CT volumes. Therefore, we would like to consider differences of 1) similarity of structure and 2) intensity range among two domains on the loss function.

The first loss term is based on  SSIM\cite{SSIM} (structure similarity). SSIM is an evaluation criterion of similarity of structure  between two images. We define the SSIM term for our proposed loss function by
\begin{equation}
L_{\rm S}(\textbf{\textit{x}},\textbf{\textit{y}}) = \frac{1}{N} \left\{ 1-\frac{(\mu_{\textbf{\textit{x}}}\mu_{\textbf{\textit{y}}}+C_1)(2\sigma_{\textbf{\textit{xy}}}+C_2)}{(\mu_{\textbf{\textit{x}}}^2+\mu_{\textbf{\textit{y}}}^2+C_1)(\sigma_{\textbf{\textit{x}}}^2+\sigma_{\textbf{\textit{y}}}^2+C_2)} \right\},
\end{equation}
where $\mu_{\textbf{\textit{x}}}$ is the average intensity of a given image $\textbf{\textit{x}}$, and $\sigma_{\textbf{\textit{x}}}$ is the variance of a given image $\textbf{\textit{x}}$. $\sigma_{\textbf{\textit{xy}}}$ is the covariance of given image $\textbf{\textit{x}}$ and $\textbf{\textit{y}}$. ${N}$, $C_1$ and $C_2$ are constant numbers. 

Moreover, regardless of intensity range differences among clinical and $\mu$CT volumes, the intensity of the images after SR should be kept as if the clinical CT volumes. We introduce new loss called the upsample and downsample loss terms, defined by
\begin{eqnarray}
L_{\rm U}(\textbf{\textit{y}})&=&{\rm MSE} (\textbf{\textit{y}},g(\textbf{\textit{y}}^{LR})), \\
L_{\rm D}(\textbf{\textit{x}})&=&{\rm MSE} (\textbf{\textit{x}},f(\textbf{\textit{x}}^{SR})),
\end{eqnarray} where $g$ represents the nearest-neighbor upsampling function that could rescale an image 8-times larger than its original size and $\textbf{\textit{y}}^{LR}$ is the fake clinical CT image generated by the generator $G_2$. $f$ is the average pooling function that  rescales an given image to 1/8 of its original size and $\textbf{\textit{x}}^{SR}$ is super-resolution result generated by the generator $G_1$. 
We calculate the MSE (mean squared error) inside these equations. Although this does not directly influence the SR result, it helps to maintain the intensity and structure when translating images from $\mu$CT domain to clinical CT domain. Then we translate the image back to $\mu$CT domain again.

Here, we write the overall loss function of CycleGAN as
\begin{equation}
\begin{split}
 L(G_1,G_2,D_X, D_Y)&= L_{\rm orig}(G_1, G_2, D_X, D_Y, X, Y) \\
                   &+\lambda_1 L_{\rm S}(\textbf{\textit{x}},f(\textbf{\textit{x}}^{SR})) +
                   \lambda_2 L_{\rm S}(\textbf{\textit{y}},g(\textbf{\textit{y}}^{LR}))\\
                  &+\lambda_3 L_{\rm D}(\textbf{\textit{x}},f(\textbf{\textit{x}}^{SR})) +\lambda_4 L_{\rm U}(\textbf{\textit{y}},g(\textbf{\textit{y}}^{LR})),
\end{split}
\end{equation}
\noindent where $X$ is the domain of image $\boldsymbol{x}$, $Y$ is the domain of image $\boldsymbol{y}$, $ L_{\rm orig}(G_1, G_2, D_X, D_Y, X, Y)$ is a term consisting of loss function that are used in original CycleGAN\cite{CycleGAN}. $\lambda_1$, $\lambda_2$, $\lambda_3$, $\lambda_4$ are the weights of each loss term.

\subsection{Super-resolution CycleGAN (SR-CycleGAN)}
CycleGAN can learn to translate an image from a source domain $\textit{X}$ to a target domain $\textit{Y}$ in the absence of paired examples. The mathematical idea of CycleGAN is to get an mapping $G_1$ : $\textit{X} \rightarrow \textit{Y}$ and another translator $G_2$: $\textit{Y} \rightarrow \textit{X}$. A loss term called ``cycle consistency loss" is added to encourage $G_2(G_1(\textbf{\textit{x}})) \approx \textbf{\textit{x}}$ and $G_1(G_2(\textbf{\textit{y}})) \approx \textbf{\textit{y}}$, where $\textbf{\textit{x}}$ are images from domain $X$ and $\textbf{\textit{y}}$ are images from domain $\textit{Y}$. An discriminator $D_1$ is added to classify whether a given image is definitively from domain $\textit{Y}$ or generated by the generator $G_1$ from domain $X$. Another discriminator $D_2$ is added to classify a given image is definitively from domain $\textit{X}$ or generated by the generator $G_2$ from domain $\textit{Y}$.

\subsection{Super-resolution UNIT (SR-UNIT)}

UNIT can be seen as a variantion of CycleGAN. When facing with super-resolution problem, UNIT has problems that are similar to CycleGAN: its loss function also could not meet the requirements of super-resolution problem, and it is not a SR network. We name the modified UNIT as SR-UNIT. Structure of SR-UNIT is also shown in Fig. 1.

\subsection{Super-resolution process}
Lung regions can be obtained by simple thresholding followed by morphological operation to fill holes and remove excess regions. Intensity normalization is also performed for each scanning modality.

For training, we obtain 2D patches both from clinical CT volumes and $\mu$CT volumes and use them for training SR-CycleGAN or SR-UNIT. Patch size is 32$\times$32 pixels from the clinical CT, and 256$\times$256 from the $\mu$CT. We took 2000 patches randomly from each clinical CT and $\mu$CT volumes.
For inference, we obtain the trained SR network generator $G_1$ for testing. We took test patches from clinical CT. The input is a 2D patch cropped from clinical CT volume. For reducing the texture of junction between each patches, size of test patches are larger than the training. Output patch is 8-times larger (e.g., 8$\times$8 input, 64$\times$64 output) than the input patch, which we call it the SR patch. We conjoin output patches to obtain the SR result of whole lung.

\section{Experiments and Results}

\subsection{Dataset} 
We evaluated the proposed  method on five clinical CT volumes and  five corresponding micro-CT volumes of lung cancer specimens obtained after lung resection surgeries. The clinical CT volumes were scanned by a clinical CT scanner (SOMATOM Definition Flash,  Siemens Inc., Munich, Germany). The resolution of the clinical CT volume was 0.625$\times$0.625$\times$0.6 $\mu$m. The micro CT volumes were scanned by a micro-CT scanner (inspeXio SMX-90CT Plus, Shimadzu, Kyoto, Japan). The lung cancer specimens were scanned with isotropic resolutions in the range of 42-52 $\mu$m.

\subsection {Parameter Settings} 
In the training phase, we extracted 2000 patches from each case. The size of patches extracted from clinical CT volumes were of 32$\times$32 pixels. The size of patches extracted from $\mu$CT volumes were of 256$\times$256 pixels. Since super-resolution always enlarged the images to power of 2 times, and comparing the resolution of clinical CT volumes (625$\mu$m) and $\mu$CT volumes (52$\mu$m), we considered 8-times super-resolution to be the most proper. The weights of proposed loss function were set empirically as $\lambda_1=2.0$ and $\lambda_2=\lambda_3=\lambda_4=1.0$. About parameters of SSIM loss $L_{\rm S}$, we set $N=1$, $C_1 = 0.02$ and $C_2 = 0.06$. Training epoch was 200. Number of total patches was 10000.

\begin{figure}
   \begin{center}
    			\begin{center}
    				\includegraphics[width=0.99\textwidth]{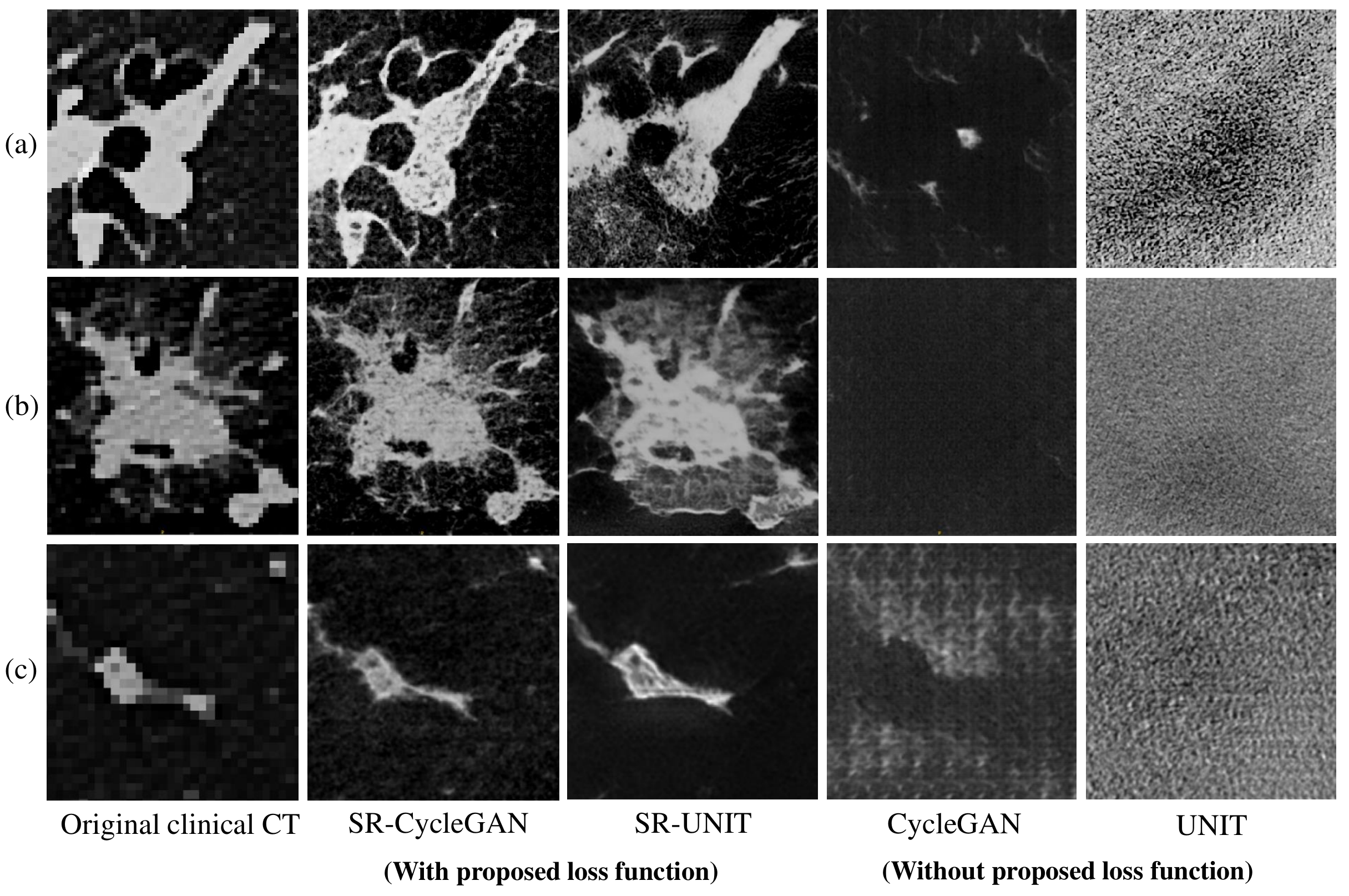}
    			\end{center}
   \end{center}
   
   \caption[example] 
   { \label{fig_Experiment_result_1} 
Comparison of SR-CycleGAN, SR-UNIT and CycleGAN, UNIT. (a) Images cropped from bronchus region. (b) Images cropped from tumor region. (c) Images cropped from vessel. We could obtain both SR-CycleGAN and SR-UNIT could perform SR of clinical CT, while SR-CycleGAN outperforms other methods, especially in bronchus region. In addition, SR-CycleGAN could rebuild the bronchus walls while SR-UNIT could not. Original CycleGAN and UNIT failed to generate SR images.}
\end{figure}

   \begin{figure}
   \begin{center}
    			\begin{center}
    				\includegraphics[width=1.0\textwidth]{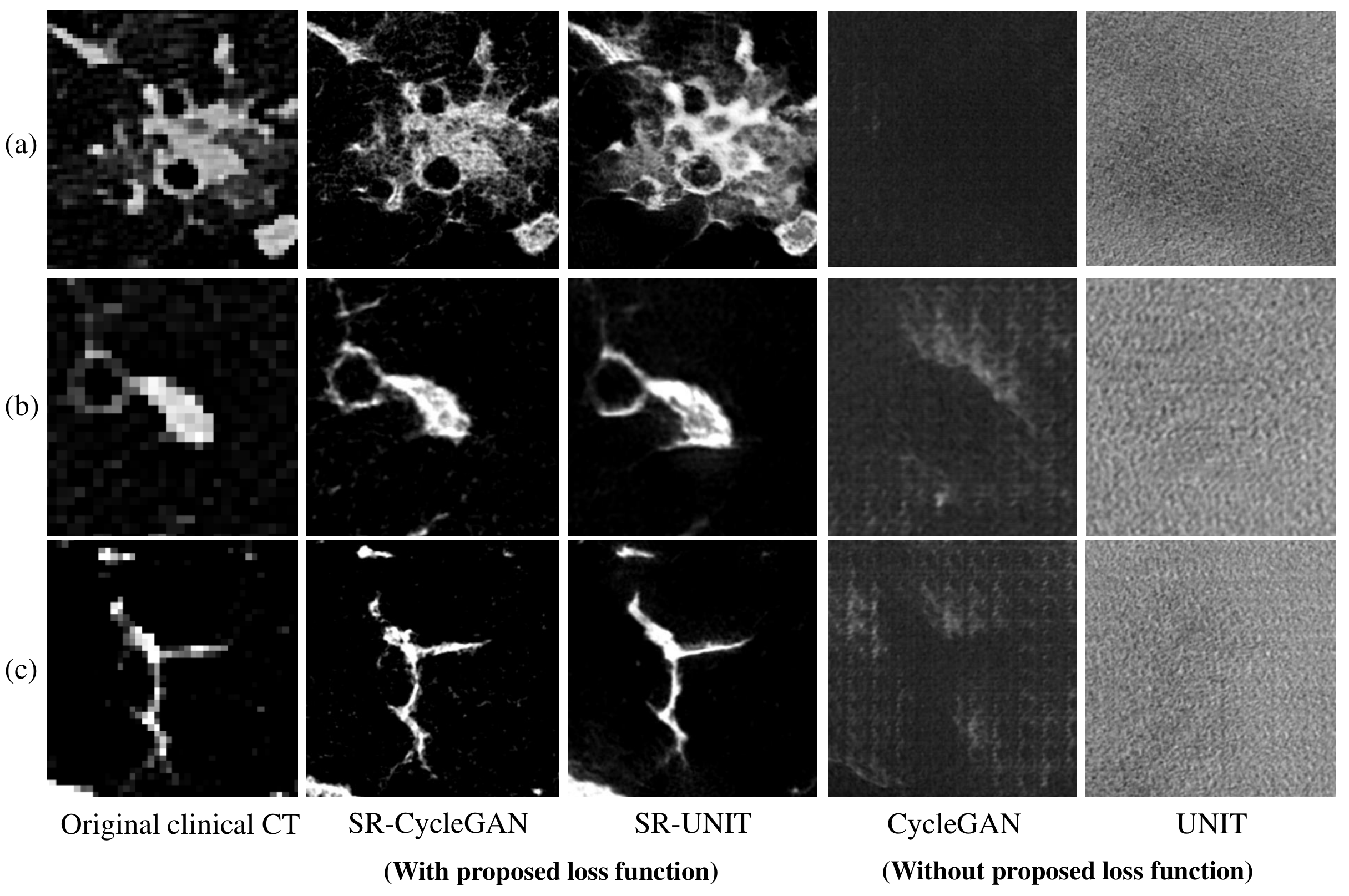}
    			\end{center}
   \end{center}
   \caption[example] 
   { \label{fig_Experiment_result_2} 
More examples of CycleGAN and UNIT with/without proposed loss function. Row (a) are images cropped from tumor region. Row (b) are images cropped from bronchus region. Row (a) are images cropped from vein region.}
   \end{figure} 

\subsection{Results}
SR results of our proposed methods were compared with original CycleGAN, original UNIT, as shown in Figs. 2 and 3.   Lung anatomies, such as the bronchus looks more clearly than bicubic-interpolation. Original CycleGAN's  and UNIT's result has produced very different results from original clinical CT volumes. These results demonstrate the proposed loss function works well for clinical CT image super-resolution.

\section{Discussion} 
\subsection{Comparing SR-CycleGAN and SR-UNIT}
The first thing we could obtain is that after the modification of loss function, both CycleGAN and UNIT performed SR on clinical CT successfully. We could obtain results that SR-CycleGAN almost performed better than SR-UNIT qualitatively. The pathiological information was kept after SR: in SR result of CycleGAN, small structures are such as vein and bronchus were well preserved.

One drawback is that SR result of SR-CycleGAN have artifact like that appeared in $\mu$CT, which makes it noiseable. By contrst, SR result of SR-UNIT do not have much artifact like that appeared in CycleGAN.

\subsection{Difficulty of quantitative evaluation}
Quantitative evaluation is usually conducted by comparing paired SR and original images. However, it is infeasible to obtain such pairs between clinical CT and $\mu$CT volumes, as also mentioned in Introduction. In this scheme, feasible quantitative evaluation approach is only to compare original clinical CT volumes and their SR results. This approach is possible by using some metrics like MSE (mean squared error) or PSNR (Peak signal-to-noise ratio). These metrics evaluates how our method produced similar intensities to the original clinical CT volumes without destroying intensity distribution or appearance structures.  
However, we also believe that this approach is still not complete as quantitative evaluation. Finding ways for quantitative evaluation is our future work. 

\section{Conclusions and Future Work}

\subsection{Conclusions}
Newly proposed loss function named MMSR loss were added to CycleGAN and UNIT for maintaining image structure and intensity, as well as avoiding generate arbitrary images after SR.  Image translation generators of the networks were replaced by image SR generators as well. Proposed methods are called SR-CycleGAN and SR-UNIT. Experiments showed proposed method successfully performed SR of lung clinical CT images into $\mu$CT level, while original CycleGAN and UNIT just produced blank images.

\subsection{Future Work}
Future work includes quantitative evaluation of the proposed methods. Since it is infeasible to obtain paired HR- and LR-data, we could not evaluated the similarity such as PSNR and SSIM directly. Furthermore, although the proposed methods focused on SR of clinical CT to $\mu$CT scale, the method is not specific to lung clinical CT SR task. It could be applied to other SR task using medical images as processing target, such as SR of $\mu$CT into H\&E-stained image scale. Since it is often difficult to register images from modalities with different resolutions, we believe that SR methods with training by unpaired LR- and HR- images will be important and widely used in the near future.

\acknowledgments 
Parts of this research was supported by MEXT/JSPS KAKENHI (26108006, 17H00867 and 17K20099), the JSPS Bilateral International Collaboration Grants, the AMED (18lk1010028s0401 and 19lk1010036h0001) and the Hori Sciences \& Arts Foundation. 


\bibliography{report}   
\bibliographystyle{spiebib}   

\end{document}